# Automated HER2 Scoring in Breast Cancer Images Using Deep Learning and Pyramid Sampling

Sahan Yoruc Selcuk[1,2,3], Xilin Yang[1,2,3], Bijie Bai[1,2,3], Yijie Zhang[1,2,3], Yuzhu Li[1,2,3], Musa Aydin[1,2,3], Aras Firat Unal[1,2,3], Aditya Gomatam[1,2,3], Zhen Guo[1,2,3], Darrow Morgan Angus[4], Goren Kolodney[5], Karine Atlan[6], Tal Keidar Haran[6], Nir Pillar[1,2,3] and Aydogan Ozcan[1,2,3,7]

[1]Electrical and Computer Engineering Department, University of California, Los Angeles, CA, USA

[2]Bioengineering Department, University of California, Los Angeles, CA, USA

[3]California NanoSystems Institute, University of California, Los Angeles, CA, USA

[4]Department of Pathology and Laboratory Medicine, University of California at Davis, Sacramento, CA, USA

[5]Bnai-Zion Medical Center, Haifa, Israel

[6]Hadassah Hebrew University Medical Center, Jerusalem, Israel

[7]David Geffen School of Medicine, University of California, Los Angeles, CA, USA

**Abstract:**

Human epidermal growth factor receptor 2 (HER2) is a critical protein in cancer cell growth that signifies the aggressiveness of breast cancer (BC) and helps predict its prognosis. Accurate assessment of immunohistochemically (IHC) stained tissue slides for HER2 expression levels is essential for both treatment guidance and understanding of cancer mechanisms. Nevertheless, the traditional workflow of manual examination by board-certified pathologists encounters challenges, including inter- and intra-observer inconsistency and extended turnaround times. Here, we introduce a deep learning-based approach utilizing pyramid sampling for the automated classification of HER2 status in IHC-stained BC tissue images. Our approach analyzes morphological features at various spatial scales, efficiently managing the computational load and facilitating a detailed examination of cellular and larger-scale tissue-level details. This method addresses the tissue heterogeneity of HER2 expression by providing a comprehensive view, leading to a blind testing classification accuracy of 84.70%, on a dataset of 523 core images from tissue microarrays. Our automated system, proving reliable as an adjunct pathology tool, has the potential to enhance diagnostic precision and evaluation speed, and might significantly impact cancer treatment planning.

## Introduction

Breast cancer (BC) is one of the most common types of cancer globally, ranking as the most prevalent cancer among women (excluding non-melanoma skin cancers) and the second-leading cause of cancer-related deaths among women after lung cancer (*1*, *2*). The complex and varied nature of BC necessitates accurate histological diagnostic procedures, such as determining the status of the human epidermal growth factor receptor 2 (HER2) (*3*). HER2 protein plays a significant role in the growth of cancer cells and is a key indicator of BC aggressiveness. The level of HER2 protein expression has prognostic and predictive value in determining patient outcomes (*4*).

In clinical practice, the assessment of HER2 status is mostly conducted through immunohistochemical (IHC) staining (*5*), followed by manual inspection of tissue slides by certified pathologists, a process depicted in Fig. 1. The American Society of Clinical Oncology/College of American Pathologists (ASCO/CAP) guidelines, published in 2018 (*6*) and affirmed in 2023 (*7*), outline specific scoring criteria for this assessment. A tumor is considered HER2 positive if it shows strong, complete, and intense membrane staining (3+) in more than 10% of tumor cells. If the staining is weak to moderate and complete in more than 10% of tumor cells, the case is scored as equivocal (2+). Cases with no staining or incomplete, barely perceptible membrane staining in 10% of tumor cells or less are classified as HER2 negative (0+ or 1+).

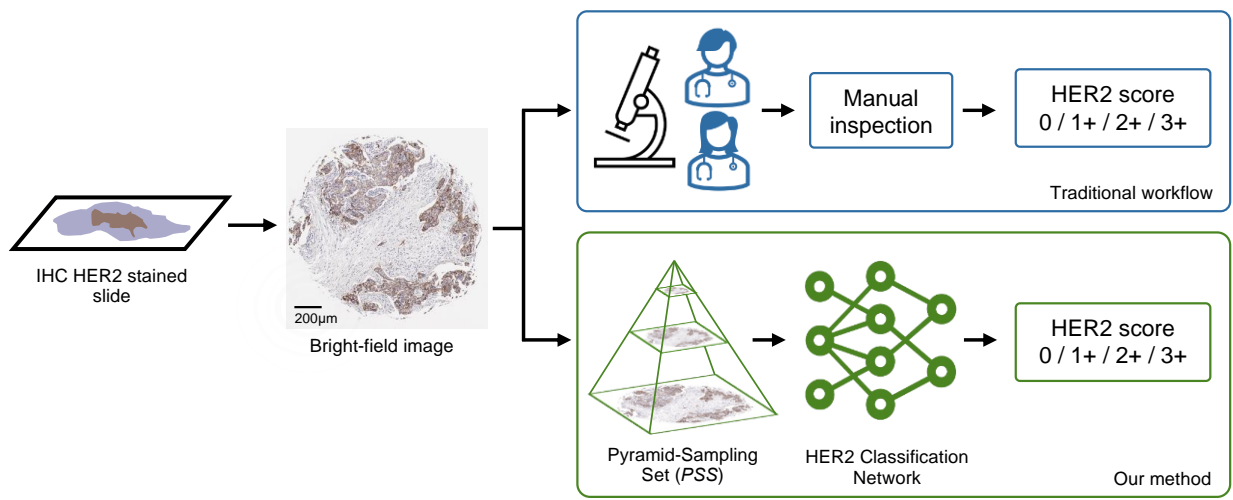

**Figure 1. Comparison of traditional HER2 scoring and the presented deep learning-based method.** The traditional HER2 score evaluation depends on manual inspection of tissue slides by pathologists. Our presented methodology introduces automated HER2 scoring using pyramid sampling and a classification neural network.

Although this traditional method is widely adopted, several challenges emerge in the manual evaluation of IHC slides. Reproducibility and concordance among pathologists are poor, and this may compromise diagnostic accuracy (*8*, *9*). Additionally, this manual evaluation process is notably time-consuming and requires careful examination by pathologists. These challenges are further exacerbated in resource-constrained areas, where the availability of expert breast pathologists may be limited, making HER2 assessment even more challenging (*10*, *11*).

Therefore, there is a growing need for automated tools that can assist pathologists in the evaluation of IHC images (*12*) to streamline the assessment of HER2 status, enhancing its efficiency, consistency and reliability (*13*, *14*). Such tools need to reduce the time required for analysis, decrease false positives and negatives, and reduce variability in measurements and require rigorous test sets for evaluation (*15*). However, due to the lack of high-quality large public datasets with rigorous established labels, most of the demonstrated approaches are only assessed on small, hand-picked image patches which failed to capture the tissue heterogeneity and sample variations typically encountered in clinical settings. A notable contest (*16*) entailed 86 whole-slide images (WSIs) where only 28 were selected for testing. Strong discordance between experts and clinical reports was observed with most discrepancy falling between HER2 1+ and 2+. Nonetheless, automated HER2 scoring systems have seen significant advances through various computational methodologies. Early research in this domain utilized standard image processing techniques and machine learning methods. Further work investigated the use of local binary patterns (LBP) and color features in conjunction with machine learning algorithms. Leveraging intensity and color features alongside uniform LBP, Singh *et al.* achieved a 91.1% accuracy using a neural network classifier on a filtered set of 371 image patches, specifically excluding outliers and ensuring a minimum of 80% content of interest within each patch (*17*); successive research by the same group deployed characteristic curves and uniform LBP features with logistic regression and support vector machine (SVM) classifiers for HER2 score assessment, while omitting image tiles with less than 40% of the region of interest (ROI) (*18*).

The adoption of deep neural networks (DNNs), especially convolutional neural networks (CNNs) marks a recent shift in computational pathology, which is largely driven by neural networks' ability to analyze and interpret complex patterns in histopathology images (*19*). For instance, CNNs processing 128×128-pixel HER2 image patches for score classification achieved a 97.7% accuracy on 119 core regions from 81 WSIs by manually selecting small, reliable regions for classification (*20*). Combining SVM, random forest, and a CNN for HER2 scoring with color deconvolution and watershed segmentation resulted in an accuracy of 83% (*21*). Additionally, fully connected long short-term memory (LSTM) networks have been used for segmenting and labeling cell membranes and nuclei in HER2-stained tissue samples, reaching 98.33% accuracy on a set of 752 image patches from 79 WSIs, highlighting a selective analysis with data exclusion (*22*). Other approaches, including deep reinforcement learning for ROI-based score prediction, and a modified U-Net architecture for WSI segmentation and tissue classification, have demonstrated HER2 classification accuracies of 79.4% and 87% over datasets of 86 and 127 WSIs, respectively (*23*, *24*). In addition to these, there has been increasing interest in exploring multiple instance learning (MIL) methods to enhance the analysis of histopathological images. MIL is a form of weakly supervised learning where training instances are arranged in sets, called bags, and a label is provided for the entire bag – usually on WSI level – instead of individual instances/image patches (*25*). Liu *et al.* implemented a MIL-based weakly supervised learning framework, evaluated on a dataset of 251 slides and achieving a 55% accuracy in classifying HER2 (*26*).

The preceding studies have delved into automating HER2 scoring employing diverse techniques, ranging from image processing to advanced machine learning techniques, predominantly using patches from a single resolution level, neglecting features observable at the broader tissue context, which is essential for precise HER2 evaluation. Additionally, investigations employing MIL necessitate an exhaustive analysis of every potential high-resolution patch within WSIs, resulting in a heavy computational load. Furthermore, most of the existing methods preselect small ROIs from WSIs or tissue cores in their training and testing. Such sampling strategies underrepresent tissue complexity and variability, potentially leading to overestimated performance metrics and a lack of generalizability.

In this work, we introduce an automated, deep learning (DL)-based HER2 score classification framework, illustrated in Fig. 1. Contrasting the aforementioned approaches, our method is based on a pyramid sampling strategy and a HER2 score inference protocol (as shown in Figs. 2a and 2d), addressing the classification challenge of HER2 expression heterogeneity. By using a randomly selected subset of high-resolution patches rather than exhaustively analyzing all possible ones, we significantly enhance computational efficiency without compromising HER2 score inference accuracy. Moving beyond conventional single-resolution-based image analyses, our pyramid sampling framework integrates detailed cellular features with broader tissue architecture, offering a comprehensive representation of HER2 expression patterns and a complete perspective on tissue heterogeneity. Our inference protocol analyzes morphological features at multiple spatial scales, efficiently balancing detailed cellular analysis with broader tissue examination. This approach, which captures and integrates patches of varying scales from high-resolution images into a Pyramid-Sampling Set (*PSS*) for DL-based evaluation, not only tackles the issue of HER2 expression heterogeneity but also achieves an accuracy of 84.70% in blind testing across 523 tissue core images obtained from 300 patients. This automated approach can help standardize HER2 assessment, streamline pathologists' workflow, and improve diagnostic accuracy.

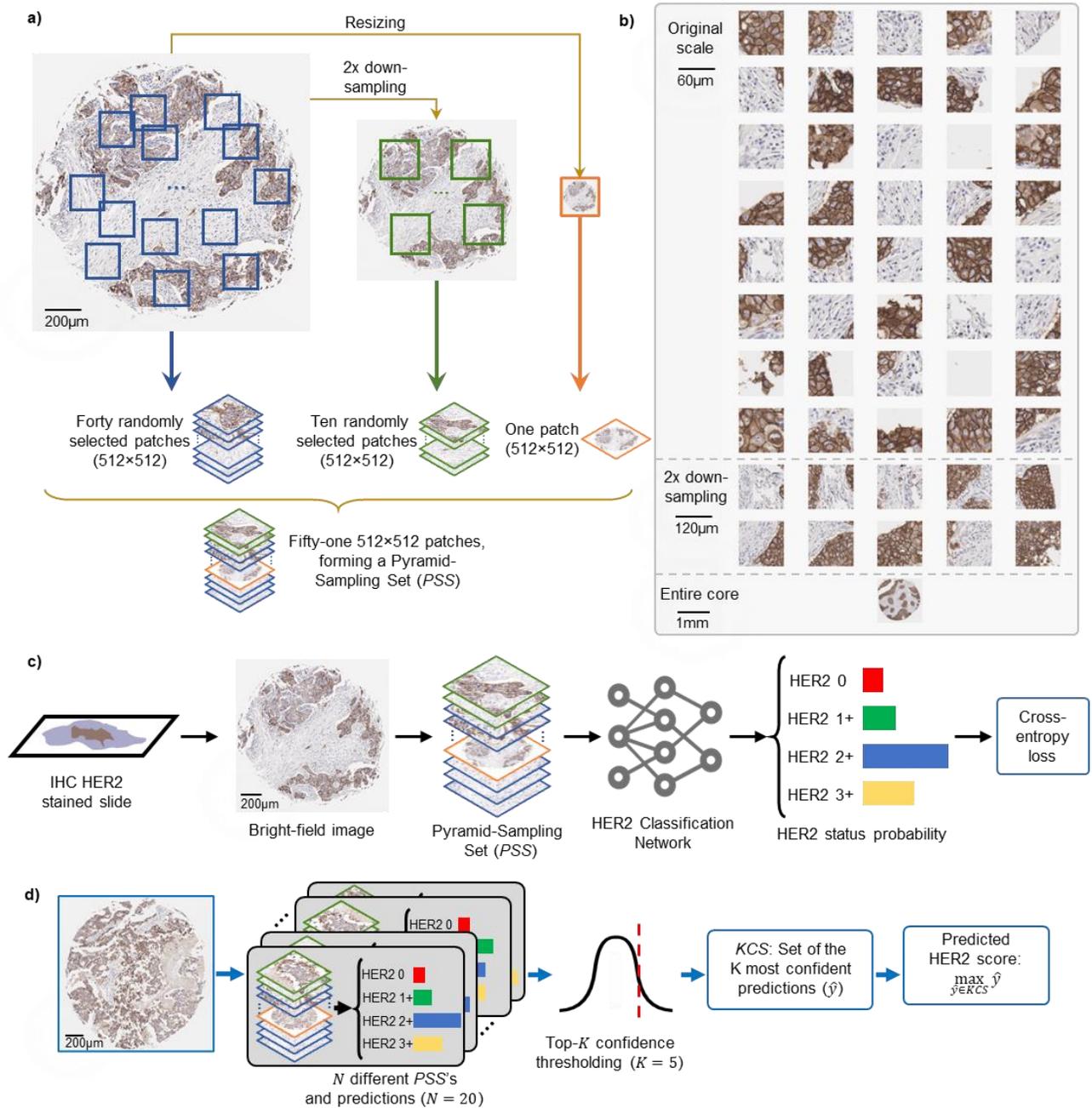

**Figure 2. Overview of the automated HER2 score classification framework. a)** Formation of a Pyramid-Sampling Set (*PSS*), detailing the extraction of multi-resolution patches from tissue images. **b)** An instance of a *PSS*, showing 40 randomly selected patches from the original resolution, 10 randomly selected patches from the half resolution case (2x-downsampled) and the entire core image resized to the patch resolution. **c)** The training process of the presented methodology involving pyramid sampling and backpropagation through a deep network. **d)** The presented inference protocol, entailing the formation of multiple *PSS*s and the final HER2 score prediction as the maximum HER2 score among the predictions with top-*k* confidence.

# Results

Our study introduces a DL-based HER2 classification method utilizing *PSS*s to analyze BC tissue samples. The classification network was trained on a dataset of 1462 core images from 823 patients, with an additional set of 162 cores from 149 patients used for validation. The efficacy of the model was blindly evaluated using a set of 523 core images from 300 patients that were not previously seen by the model during the training or validation phases. We evaluated the performance of our model with both qualitative and quantitative analyses. While creating the ground truth of our dataset, we involved five board-certified pathologists to independently score the cores. This approach mitigated the risk of relying on possibly inaccurate patient records due to tissue heterogeneity, providing a robust methodology for dataset labeling that prioritizes precision in reflecting real-world diagnostic scenarios. We introduced a voting protocol (detailed in the Methods section) to resolve inter-pathologist variability and to achieve consensus in core evaluations. Non-diagnostic cores were excluded to ensure the high quality and diagnostic relevance of the compiled dataset. Overall, we compiled a comprehensive and accurately labeled dataset of 2147 cores from 1272 patients.

Our model's training and testing protocols are detailed in Figs. 2c-d. During training, each core image is transformed into a *PSS* and fed into the DL model with its ground truth label. We compute a cross-entropy loss to optimize the model, which upon convergence, enters the testing phase. In testing, for each core image, $N$ independent *PSS*s are generated. We then conduct a forward process on these sets, selecting the top $k$ predictions with the highest confidence, forming the $k$-Confident Selection Set (*KCS*). The final score is the highest score within the *KCS*, effectively addressing the heterogeneity of HER2 expression. Therefore, the two key hyperparameters in this process are $N$, the number of independent *PSS*s generated per tissue core, and $k$, the number of high-confidence predictions used for final scoring. This method ensures a diverse representation of HER2 status and focuses on the most confident predictions to enhance HER2 scoring accuracy. By prioritizing predictions with the highest confidence, our approach reduces the chance of inaccuracies due to lower-confidence inferences, ensuring the final HER2 score is based on significant expressions. This methodology not only improves the reliability of HER2 scoring but also captures essential expressions critical for a comprehensive assessment of HER2 status in breast tissue sections.

We demonstrate the capability of our automated HER2 scoring system with 12 examples in Fig. 3. This figure illustrates the distribution of the predicted HER2 scores coming from the most confident five predictions for a subset of the test samples, which were accurately classified into the four HER2 categories. Each sample underwent evaluation by generating 20 independent *PSS* predictions, with the subsequent histograms depicting the score distributions from the five highest-confidence *PSS*s. These predictions were then grouped into color-coded categories corresponding to the consensus HER2 score: 0, 1+, 2+, and 3+.

A key observation from Fig. 3 is the influence of HER2 expression heterogeneity on the predictions of the *PSS*s with the highest confidence. For example, within the yellow-coded box highlighting HER2 3+ samples, there is a noticeable variation in the intensity of HER2 biomarker expression. The last core image within this grouping shows a pronounced level of HER2 positivity, which is consistently recognized across all five high-confidence *PSS* predictions as 3+. In contrast, the first core image of the same category exhibits a lower intensity of HER2 expression, leading to a slight variance where two out of five high-confidence *PSS*s predict a 2+ score. Similarly, the green and blue boxes corresponding to HER2 1+ and HER2 2+ categories, respectively, also demonstrate this trend. The model's predictions reflect the level of HER2 expression, with the majority of high-confidence *PSS*s aligning with the consensus category in most samples. However, some *PSS*s indicate adjacent categories, suggesting a borderline expression level. The red box, delineating HER2 0 samples, is particularly noteworthy, as all high-confidence *PSS*s consistently predict a HER2 score of 0.

Monte Carlo simulations were also leveraged to reveal the effects of varying the number of independent *PSS*s. For each sample, we employed our converged model and tested samples with varying $N$ and $k$ values, as well as different sets of *PSS*s, to validate the stability and consistency of our model; refer to the Methods section for details. Classification accuracy was markedly improved as the number of *PSS*s increased up to a certain point, with a fixed confidence threshold parameter of $k=5$ as shown in Fig. 4. A notable peak in accuracy is observed when $N$ is set to 20, where the maximum classification accuracy reaches 87.76%. As we continued to increase $N$ to 200, the accuracy gain became marginal. Confusion matrices corresponding to the minimum, median, and maximum accuracy benchmarks, achieved with $N=20$ and $k=5$, provide a quantitative view of the system's performance as shown in Fig. 4. Specifically, the minimum accuracy recorded is 82.52%, the median accuracy stands at 84.70%, and the maximum accuracy reaches 87.76%. Initially, at lower values of $N$, there is a wider spread between these accuracy measures, indicating variability in the model's performance. This implies that to decrease the variation in predictions caused by random sampling, employing a larger $N$ is effective at the cost of testing speed. In doing so, the performance is expected to align with the median value observed under the default configuration ($N = 20, k = 5$).

Next, we focused on the precision of classification accuracy across varying confidence threshold parameters, denoted by the parameter $k$, while holding the number ($N$) of independent *PSS*s used in the inference protocol constant at 200, shown in Fig. 5. This figure highlights the delicate balance between the confidence in prediction and the precision of the final score, and it illustrates the performance metrics at different accuracy levels using confusion matrices. The accuracy trends depicted in the graph provide an illustration of the model's classification performance against varying $k$ values. Notably, when the confidence threshold parameter ($k$) value exceeded 20, there was a pronounced decrement in the model's accuracy. This was consistently observed across the minimum, median, and maximum accuracy rates. This trend suggests that a *k* value within the range of 1 to 20 maintains optimal classification performance, whereas higher *k* values lead to a marked reduction in accuracy.

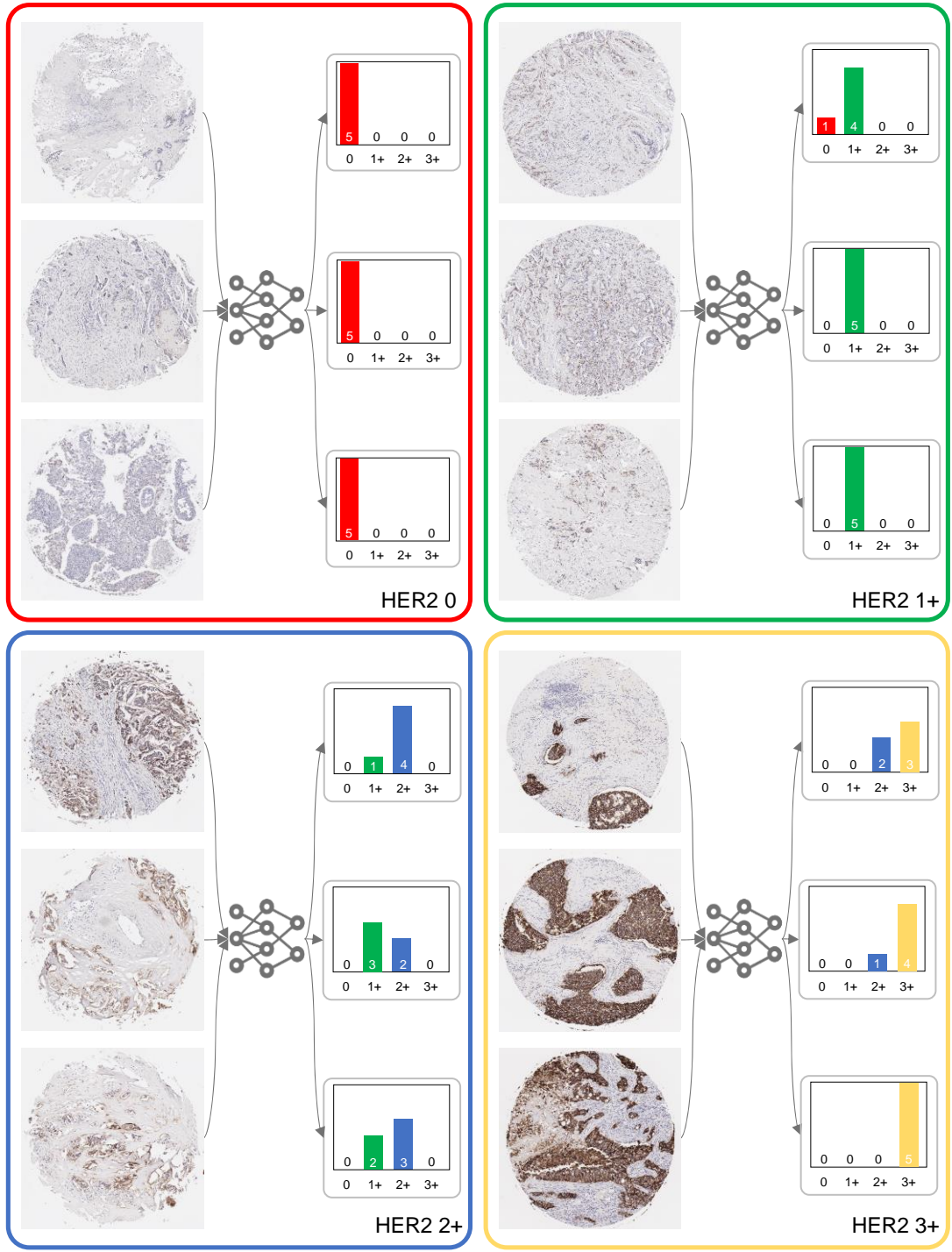

**Figure 3. Distribution of the predicted HER2 scores for 12 randomly selected test samples, as determined by the *PSS* approach.** For each sample, *N*=20 independent *PSS* predictions are generated, and the histograms display

the HER2 score distributions from the $k=5$ *PSS*s with the highest confidence levels. The final HER2 score prediction for each sample is generated by the maximum score from these top-5 confidence *PSS*s. Samples are grouped and color-coded according to their consensus HER2 score categories: 0, 1+, 2+, and 3+.

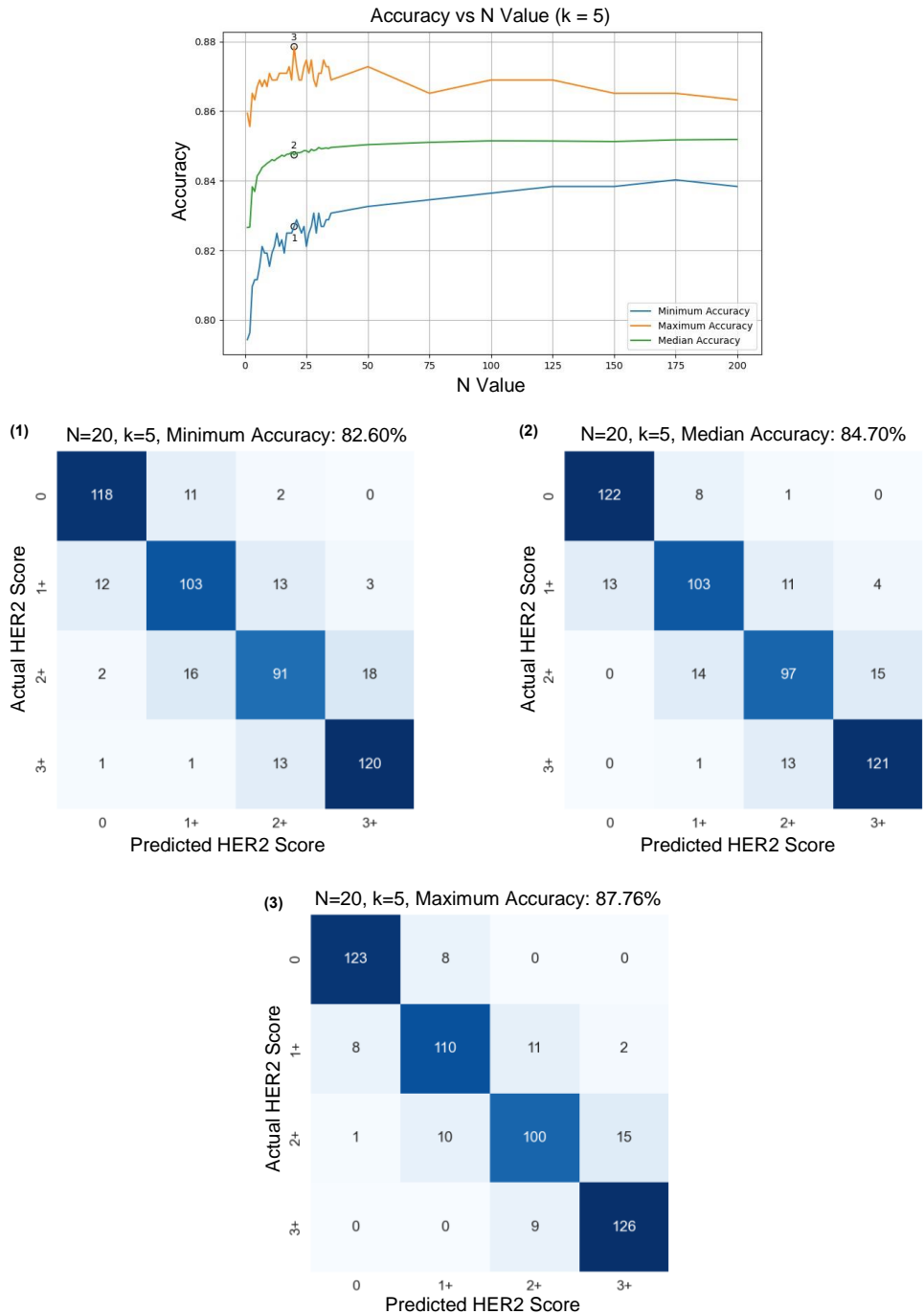

**Figure 4. Relationship between the number of independent *PSS*s and HER2 scoring accuracy.** The top plot displays how accuracy changes as a function of *N* used during inference, with a fixed confidence threshold parameter ($k=5$). The accompanying confusion matrices exemplify the accuracy at distinct performance benchmarks—minimum, median, and maximum—achieved with $N=20$ *PSS*s. These benchmarks were established through a Monte Carlo

simulation designed to evaluate the influence of *PSS* selection randomness on the overall accuracy of the HER2 score classification system. Blind testing set includes 523 core images from 300 patients that were not previously seen by the model during the training or validation.

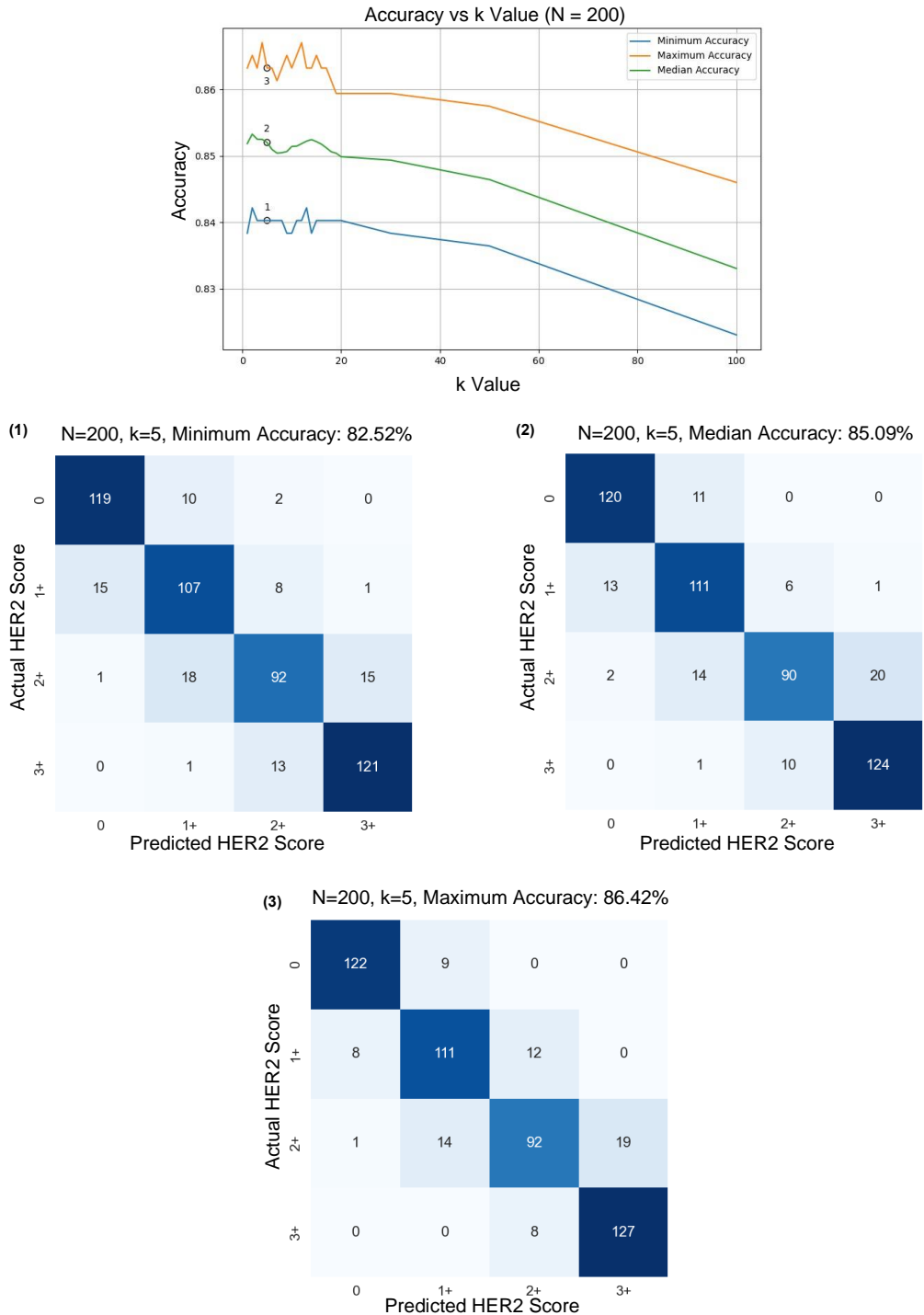

**Figure 5. Analysis of HER2 scoring precision as a function of *k* with a fixed number of independent *PSS*s (*N*=200).** The graph traces accuracy variations across different *k* values, while the confusion matrices document the system's performance at minimum, median, and maximum accuracy values for *k*=5. These analyses examine the impact of the confidence threshold parameter ($k$) on the overall accuracy of HER2 score predictions. Blind testing set includes 523 core images from 300 patients that were not previously seen by the model during the training or validation.

## Discussion

In this study, we introduced a DL-based method that utilizes pyramid sampling to automate the classification of HER2 status in IHC-stained tissue images. By leveraging a hierarchical approach that intricately analyzes features across multiple spatial scales, our method addresses the challenge of HER2 expression heterogeneity, without ROI selection prior to model training. The success of our approach is substantiated through quantitative analysis involving 523 core images from 300 patients never seen before in the training or validation, achieving a classification accuracy of 84.70% compared to the consensus scores obtained from 5 board certified pathologists. This robust performance not only underscores the method's precision in captured details but also its potential to mitigate the challenges currently faced in clinical and research settings, such as observer inconsistency and protracted diagnostic timelines.

The pyramid sampling strategy and inference protocol introduced in our study represent a significant advancement in the automated classification of HER2 status, marking, to our knowledge, the first instance of utilizing multi-scale feature analysis for automated HER2 scoring in IHC-stained tissue images. Our pyramid sampling strategy marks a significant departure from previously described methods, focusing on the detailed analysis of both membranous features at the cellular level and broader tissue regions. While most of the previous HER2 classification tools in the literature rely solely on localized patches from a single-scale WSI image (*16–18*, *20–22*), we combined high-resolution image patches with their lower-resolution counterparts, ensuring that both the micro-environment of cellular features and the macro-context of tissue architecture are captured and analyzed through our digital framework. This results in a balanced presentation of both the high spatial frequency details and a comprehensive sample field-of-view in the input of our classification network, which is important to mitigate the HER2 expression heterogeneity observed in tissue samples. This ability of our approach to mitigate the heterogeneity of HER2 expression is also exemplified in Figs. 6 and 7. These figures collectively emphasize the importance of employing multiple *PSS*s covering different spatial scales to navigate the complexities of HER2 scoring, showcasing the system's adeptness at identifying subtle differences in HER2 biomarker expression levels within the same score category. Furthermore, our method employs a random patch selection mechanism, choosing a specific number of patches at each iteration. This strategy significantly reduces the computational load, enhancing the efficiency of our method without sacrificing the quality and accuracy of tissue characterization.

Beyond the model's overall accuracy, it is informative to examine its performance in distinguishing between adjacent HER2 classes. The accuracy figures calculated for distinguishing between adjacent HER2 categories demonstrate the model's effectiveness and are suggestive of its practical utility in a clinical setting. For example, our model achieves an accuracy of 89.62% for differentiating between 0 and 1+ HER2 scores. This is clinically significant because a score of 0 will follow a treatment regimen without using HER2-targeted therapies, while a score of 1+ might prompt further analysis according to recent ASCO/CAP data (*7*). A high accuracy in this range minimizes the risk of patients being incorrectly excluded from receiving HER2-targeted treatments if they might benefit from them. For the 1+ vs 2+ categories, the model shows an accuracy of 84.68%. HER2 2+ cases often require additional confirmatory tests such as fluorescence in situ hybridization (FISH) to make a final treatment decision (*27*, *28*). A high accuracy in discriminating HER2 1+ and 2+ helps ensure that patients are appropriately triaged for FISH testing while avoiding unnecessary procedures for others. Finally, the model's accuracy in distinguishing between 2+ and 3+ scores is 87.26%, ensuring that patients with strong HER2 positivity are promptly identified for appropriate therapeutic intervention.

The practical implications of our DL-based approach for HER2 status classification touch upon two pressing issues in pathological assessment: the consistency of manual evaluations and the efficiency of the diagnostic process. Manual evaluation of HER2 IHC status is susceptible to a degree of subjectivity inherent in pathologist judgment. Assessment of data from the CAP surveys demonstrated poor agreement in the evaluation of 0 and 1+ cases (26% concordance) and 58% concordance between 2+ and 3+ (*29*). Such vast differences in HER2 quantification can have significant implications on treatment decisions and ultimately patient outcomes. By algorithmically standardizing the scoring process, our method introduces a level of consistency unattainable through manual inspection. This consistency allows our presented model to serve as a valuable enhancement tool for pathologists, offering a consistent second opinion free from human fatigue, level of experience or bias. This is especially critical in low volume pathology departments that may lack specialized breast pathology experts. In such settings, our automated method can function as a stand-in consultant, providing assessments that can be trusted to align with what a specialist might determine.

The second substantial advantage of our system lies in its potential to significantly shorten the diagnostic turnaround time. Delays in diagnosis can have detrimental effects on patient care, inducing patient dissatisfaction (*30*), anxiety, and stress (*31*). Our approach radically shortens the assessment time to seconds per case, a reduction that can have far-reaching implications in the clinical setting, especially in diagnostically challenging cases. Moreover, in busy pathology departments where the volume of cases can be overwhelming, the efficiency of our method could prevent bottlenecks and reduce inter-pathologists' consultations. By allowing for quicker throughput of cases, pathologists can allocate more time to complex cases where human expertise is indispensable.

Monte Carlo simulations were essential for addressing the statistical variability introduced by our pyramid sampling strategy and optimizing the parameters *N* and *k*, enabling us to systematically assess and mitigate the effects of randomness on the model's performance, ensuring a balanced representation of HER2 expression across tissue samples. By identifying an optimal *N*, we made sure that the model could consistently predict HER2 scores despite the variability in tissue representations. Similarly, optimizing *k* allowed us to focus on the most reliable predictions, enhancing the accuracy and reliability of the final HER2 score. The optimal balance for our model's operational efficiency occurred at *N*=20, as demonstrated in Figure 4, highlighting the importance of balancing the number of *PSS*s and model performance to avoid unnecessary computational costs, crucial for clinical use. Beyond *N*=20, adding more *PSS*s to the inference protocol did not significantly improve accuracy but maintained prediction consistency. Figure 5 shows how the HER2 scoring accuracy decreased when the inference hyperparameter *k* exceeded 20, emphasizing the need to keep *k* within a certain range for optimal classification. This result indicated the delicate balance required between the number of high-confidence predictions and their quality, with a higher *k* value increasing the risk of including less accurate predictions and potentially leading to more false positives.

In conclusion, our study represents a robust, efficient solution to the challenges of HER2 classification in BC diagnostics. By effectively leveraging pyramid sampling to address the heterogeneity of HER2 expression and demonstrating the potential to significantly streamline diagnostic processes, our approach enhances the accuracy and reliability of HER2 score classification. This research paves the way for more nuanced, faster, and more accessible diagnostics, ultimately contributing to the advancement of personalized medicine and improving patient care in oncology.

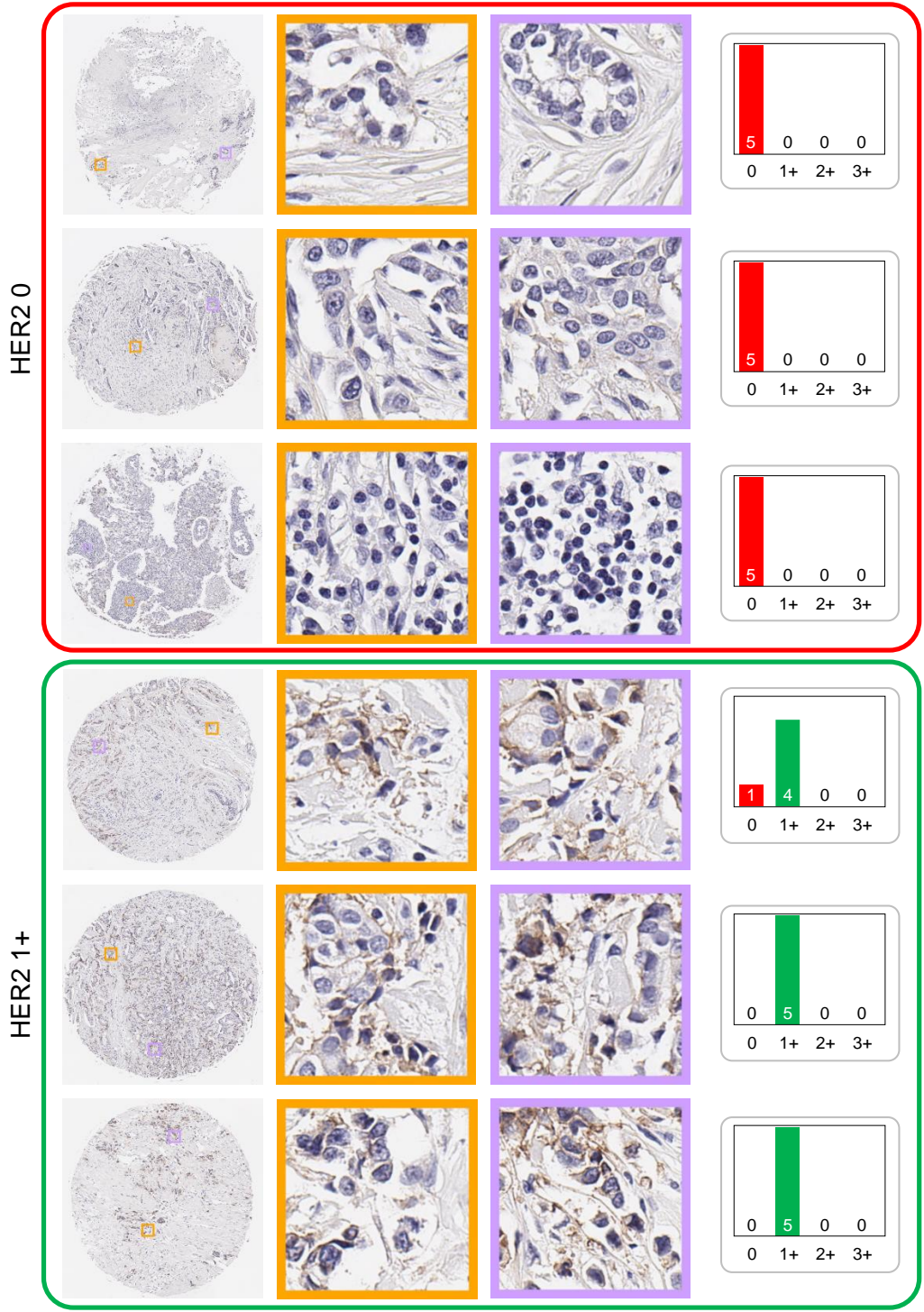

**Figure 6. Microscopic examination of HER2 scoring regions in tissue samples.** Color-coded boxes display core images from HER2 0 and HER2 1+ tissue sample categories with two magnified patches to showcase the specific histological details. Accompanying histograms indicate the predicted score distribution for the highest confidence *PSS*s, offering insights into our automated HER2 assessment.

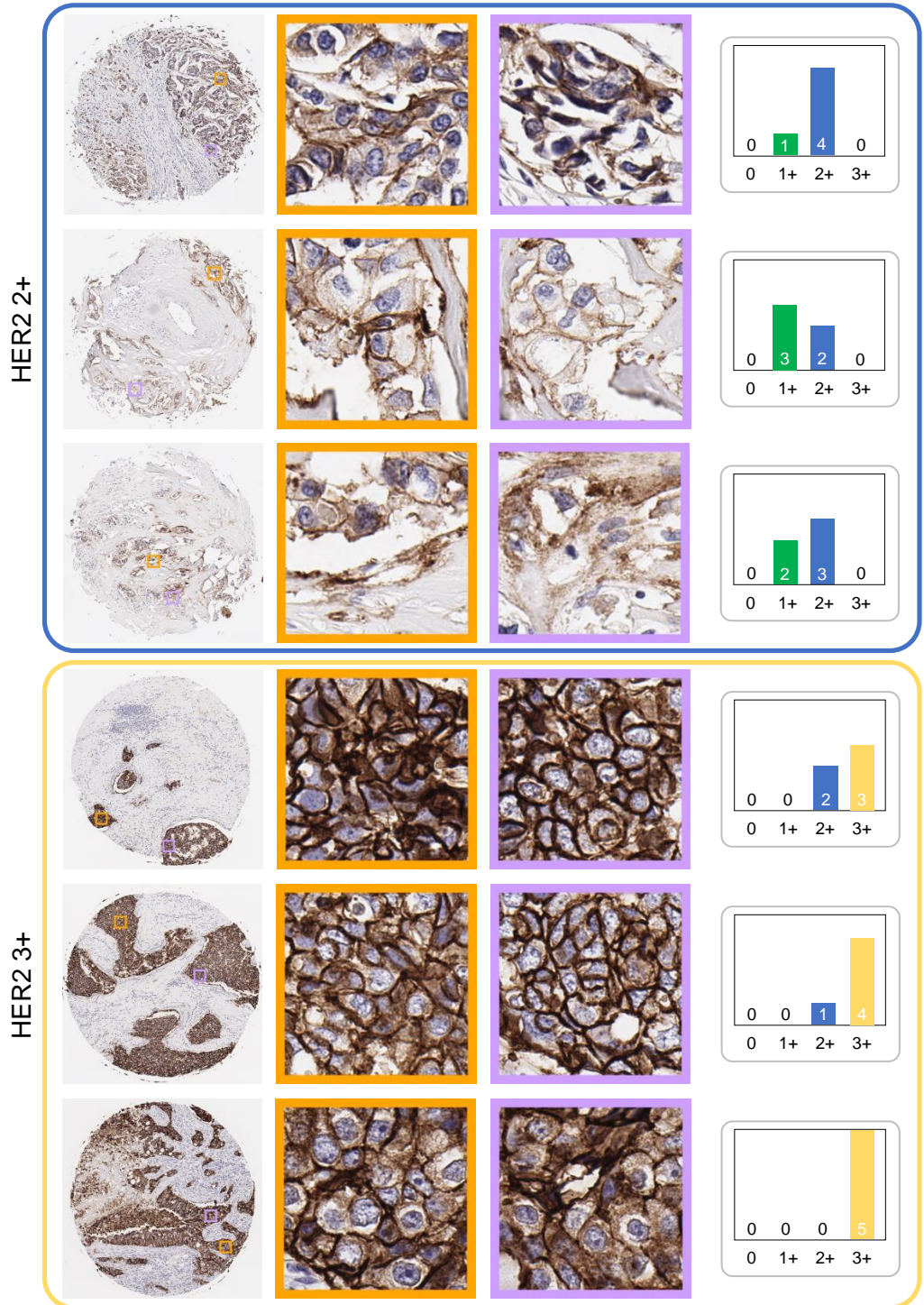

**Figure 7. Same as Figure 6, except for higher HER2 score categories.** Color-coded boxes correspond to samples within the HER2 2+ and HER2 3+ tissue sample categories. Two detailed patches are extracted from each core image, emphasizing the staining intensity and cellular patterns critical for determining the respective HER2 scores. Histograms adjacent to each sample reflect the predicted score distribution from the most confident *PSS*s.

## Methods

### Sample preparation, data acquisition and dataset creation

Fifteen unlabeled breast tissue microarray slides, each containing about 100-200 cores, were acquired from TissueArray (*32*). These samples underwent HER2 IHC staining at the UCLA Translational Pathology Core Laboratory. We captured bright-field images using a slide scanner microscope (AxioScan Z1, Zeiss) with a ×20/0.8NA objective lens (Plan-Apo). WSIs were processed to identify and extract each core, utilizing a customized algorithm based on Hough transform techniques (*33*). High-resolution bright-field images of the cores were uploaded to Google Photos for pathologist evaluations. To ensure accurate dataset labeling, we employed a voting protocol among five board-certified pathologists, where a core's evaluation needed agreement between at least two pathologists. For discordant assessments, an additional pathologist was consulted for a decisive score, leading to the core's inclusion or exclusion from the dataset. Cores of non-diagnostic quality or those deemed non-scorable were removed. We compiled 2147 cores with finalized labels, which were allocated into 1462 for the training set, 162 for validation, and 523 for testing.

### Pyramid sampling strategy

We devised a pyramid sampling strategy to capture the multi-scale nature of tissue morphology and HER2 expression patterns. This approach involved systematically extracting small patches from original high-resolution tissue images, approximately ~10,000×10,000 pixels in size. We selected forty 512x512 pixel patches directly from the original high-resolution image and ten patches from a 2x-downsampled image, alongside a single patch representing the entire tissue core resized to 512x512 pixels. These patches were combined along the channel dimension to create a *PSS*, serving as the input for our DL framework. Supplementary Figs. 1–5 present the five *PSS*s used in deriving the five highest-confidence predictions indicated in the histogram for the topmost tissue core within the green box (HER2 1+) in Fig. 3. Similarly, Supplementary Figs. 6–10 demonstrate the *PSS*s that generated the predictions for the uppermost tissue core within the yellow box (HER2 3+) in the same figure.

### HER2 score classification network architecture and training scheme

We selected DenseNet-201 (*34*) for our study, a 201-layer densely connected convolutional network, due to its efficiency in image classification. The architecture features dense blocks linked in a feed-forward manner, where each layer receives inputs from all preceding layers and forwards its feature-maps to subsequent layers, enhancing feature propagation and reuse. To integrate our pyramid sampling strategy, the first convolutional layer of DenseNet-201 was modified to accept an input with a variable number of channels, accommodating the unique depth of the *PSS*s. Furthermore, the classification layer was customized by replacing the original classifier with a fully connected linear layer tailored to the four HER2 score categories (0, 1+, 2+, 3+), employing a softmax activation function to provide class probability distribution indicative of the network's confidence in its HER2 scoring.

Addressing the class imbalance present in our tissue dataset, we implemented a balanced-class weighted cross-entropy loss function. The loss for each class was weighted inversely proportional to its frequency in the training data, allowing for an equal focus on all classes during the learning process. For the entire training set, the loss ($L$) for a batch of size ($m$) is given by:

$$L = -\frac{1}{m}\sum_{i=1}^{m}\sum_{c=1}^{C=4} w_c\, y_{i,c} \log(p_{i,c})$$

where $C$ is the number of classes (4 in our case), $w_c$ represents the class weight, $y_{i,c}$ is a binary indicator of whether class label $c$ is the correct classification for observation $i$, and $p_{i,c}$ is the predicted probability of observation $i$ for class $c$. Weights $w_c$ help to amplify the signal from underrepresented classes, driving the model to pay more attention to these classes during training. Furthermore, we applied data augmentation techniques, including horizontal and vertical flips and rotations of 90 degrees, to each image in the training set to mitigate overfitting and enhance generalizability of the trained model. The training process of our presented methodology is illustrated in Fig. 2c.

**HER2 score prediction (inference) protocol**

Following the generation of $N$ distinct predictions for each tissue core, our protocol selects a subset of $k$ predictions, *KCS*, based on their confidence levels. This process involves analyzing softmax class probabilities to identify predictions with a clear preference for one HER2 category, indicative of high confidence. The selection criterion focuses on the disparity in HER2 class probabilities within each prediction's softmax output. The final HER2 score for a tissue core is determined by aggregating the scores within the *KCS*, specifically by taking the maximum score observed across this set. Mathematically, the final HER2 score is given by the maximum score observed across the *KCS*, denoted:

$$\text{Final HER2 Score} = \max_{\hat{y} \in KCS} \hat{y}$$

where $\hat{y}$ represents the individual HER2 score predictions generated by independent *PSS*s.

**Statistical analysis and Monte Carlo simulations**

Monte Carlo simulations were utilized to analyze our HER2 score prediction model, focusing on optimizing the number of *PSS*s used in inference ($N$) and refining the selection criteria based on confidence ($k$). These simulations aimed to find an optimal balance that improves accuracy, reliability, and computational efficiency. To evaluate the impact of the inherent variability from our pyramid sampling strategy, 300 independent *PSS*s were generated for each sample in our testing set, providing diverse views of each tissue's HER2 expression. A range of values for $N$, from 1 to 200 *PSS*s per tissue sample, with $k$ fixed at 5, were explored to identify an optimal $N$ that enhances the model's consistency and robustness. Additionally, we investigated the optimal threshold for $k$ by varying its values from 1 to 20, and selected higher values [30, 50, 100], with $N$ set to a high value of 200, to refine the selection of high-confidence predictions for the final HER2 score determination and to ensure that we could investigate the effects of a wide range of *k* values. These simulations were conducted 10,000 times for each configuration of *N* and *k* to provide a statistically robust analysis.

**Implementation details**

The classification network's training was optimized using the AdamW optimizer (*35*), an advanced variant of the Adam optimizer designed to incorporate weight decay, thereby mitigating the risk of overfitting. Training commenced with an initial learning rate set at $10^{-5}$, which was dynamically adjusted in response to changes in the validation loss, with a batch size maintained at 12. This setup allowed the network to reach convergence after approximately 60 hours of dedicated training. For a core image resolution of 10,000x10,000 pixels, the typical inference time was reduced to less than 15 seconds, achieved when the inference hyperparameter *N* was set at 20. The training and testing were conducted on a desktop computer equipped with a GeForce RTX 3090 Ti graphics processing unit, 128GB of random-access memory, and an AMD Ryzen 9 5900X central processing unit. The classification network was implemented using Python version 3.12.0 and PyTorch version 1.9.0, alongside CUDA toolkit version 11.8.

# References

1. R. H. Engel, V. G. Kaklamani, HER2-Positive Breast Cancer. *Drugs* **67**, 1329–1341 (2007).

2. B. Smolarz, A. Z. Nowak, H. Romanowicz, Breast Cancer—Epidemiology, Classification, Pathogenesis and Treatment (Review of Literature). *Cancers* **14**, 2569 (2022).

3. M. Zubair, S. Wang, N. Ali, Advanced Approaches to Breast Cancer Classification and Diagnosis. *Frontiers in Pharmacology* **11** (2021).

4. K. A. B. Goddard, S. Weinmann, K. Richert-Boe, C. Chen, J. Bulkley, C. Wax, HER2 evaluation and its impact on breast cancer treatment decisions. *Public Health Genomics* **15**, 1–10 (2011).

5. H. Nitta, B. D. Kelly, C. Allred, S. Jewell, P. Banks, E. Dennis, T. M. Grogan, The assessment of HER2 status in breast cancer: the past, the present, and the future. *Pathology International* **66**, 313–324 (2016).

6. A. C. Wolff, M. E. H. Hammond, K. H. Allison, B. E. Harvey, P. B. Mangu, J. M. Bartlett, M. Bilous, I. O. Ellis, P. Fitzgibbons, W. Hanna, Human epidermal growth factor receptor 2 testing in breast cancer: American Society of Clinical Oncology/College of American Pathologists clinical practice guideline focused update. *Archives of pathology & laboratory medicine* **142**, 1364–1382 (2018).

7. A. C. Wolff, M. R. Somerfield, M. Dowsett, M. E. H. Hammond, D. F. Hayes, L. M. McShane, T. J. Saphner, P. A. Spears, K. H. Allison, Human Epidermal Growth Factor Receptor 2 Testing in Breast Cancer: ASCO-College of American Pathologists Guideline Update. *J Clin Oncol* **41**, 3867–3872 (2023).

8. M. Lacroix-Triki, S. Mathoulin-Pelissier, J.-P. Ghnassia, G. Macgrogan, A. Vincent-Salomon, V. Brouste, M.-C. Mathieu, P. Roger, F. Bibeau, J. Jacquemier, F. Penault-Llorca, L. Arnould, High inter-observer agreement in immunohistochemical evaluation of HER-2/*neu* expression in breast cancer: A multicentre GEFPICS study. *European Journal of Cancer* **42**, 2946–2953 (2006).

9. S. Di Cataldo, E. Ficarra, A. Acquaviva, E. Macii, Automated segmentation of tissue images for computerized IHC analysis. *Computer Methods and Programs in Biomedicine* **100**, 1–15 (2010).

10. J. Ferlay, H.-R. Shin, F. Bray, D. Forman, C. Mathers, D. M. Parkin, Estimates of worldwide burden of cancer in 2008: GLOBOCAN 2008. *International Journal of Cancer* **127**, 2893–2917 (2010).

11. A. Fernandes, G. Bianchi, A. P. Feltri, M. Pérez, M. Correnti, Presence of human papillomavirus in breast cancer and its association with prognostic factors. *Ecancermedicalscience* **9**, 548 (2015).

12. L. Mulrane, E. Rexhepaj, S. Penney, J. J. Callanan, W. M. Gallagher, Automated image analysis in histopathology: a valuable tool in medical diagnostics. *Expert Review of Molecular Diagnostics* **8**, 707–725 (2008).

13. J. D. Webster, R. W. Dunstan, Whole-Slide Imaging and Automated Image Analysis: Considerations and Opportunities in the Practice of Pathology. *Vet Pathol* **51**, 211–223 (2014).

14. P. W. Hamilton, P. Bankhead, Y. Wang, R. Hutchinson, D. Kieran, D. G. McArt, J. James, M. Salto-Tellez, Digital pathology and image analysis in tissue biomarker research. *Methods* **70**, 59–73 (2014).

15. M. G. Rojo, G. Bueno, J. Slodkowska, Review of imaging solutions for integrated quantitative immunohistochemistry in the Pathology daily practice. *Folia histochemica et cytobiologica* **47**, 349–354 (2009).

16. T. Qaiser, A. Mukherjee, C. Reddy PB, S. D. Munugoti, V. Tallam, T. Pitkäaho, T. Lehtimäki, T. Naughton, M. Berseth, A. Pedraza, R. Mukundan, M. Smith, A. Bhalerao, E. Rodner, M. Simon, J. Denzler, C.-H. Huang, G. Bueno, D. Snead, I. O. Ellis, M. Ilyas, N. Rajpoot, HER2 challenge contest: a detailed assessment of automated HER2 scoring algorithms in whole slide images of breast cancer tissues. *Histopathology* **72**, 227–238 (2018).

17. P. Singh, R. Mukundan, "A Robust HER2 Neural Network Classification Algorithm Using Biomarker-Specific Feature Descriptors" in *2018 IEEE 20th International Workshop on Multimedia Signal Processing (MMSP)* (2018; https://ieeexplore.ieee.org/abstract/document/8547043), pp. 1–5.

18. R. Mukundan, Analysis of Image Feature Characteristics for Automated Scoring of HER2 in Histology Slides. *Journal of Imaging* **5**, 35 (2019).

19. P. Khosravi, E. Kazemi, M. Imielinski, O. Elemento, I. Hajirasouliha, Deep Convolutional Neural Networks Enable Discrimination of Heterogeneous Digital Pathology Images. *eBioMedicine* **27**, 317–328 (2018).

20. T. Pitkäaho, T. M. Lehtimäki, J. McDonald, T. J. Naughton, "Classifying HER2 breast cancer cell samples using deep learning" in *Proc. Irish Mach. Vis. Image Process. Conf* (2016; https://aran.library.nuigalway.ie/bitstream/handle/10379/6136/IMVIP2016Book.pdf#page=88), pp. 1–104.

21. M. E. Vandenberghe, M. L. J. Scott, P. W. Scorer, M. Söderberg, D. Balcerzak, C. Barker, Relevance of deep learning to facilitate the diagnosis of HER2 status in breast cancer. *Sci Rep* **7**, 45938 (2017).

22. M. Saha, C. Chakraborty, Her2Net: A deep framework for semantic segmentation and classification of cell membranes and nuclei in breast cancer evaluation. *IEEE Transactions on Image Processing* **27**, 2189–2200 (2018).

23. T. Qaiser, N. M. Rajpoot, Learning Where to See: A Novel Attention Model for Automated Immunohistochemical Scoring. *IEEE Transactions on Medical Imaging* **38**, 2620–2631 (2019).

24. F. D. Khameneh, S. Razavi, M. Kamasak, Automated segmentation of cell membranes to evaluate HER2 status in whole slide images using a modified deep learning network. *Computers in Biology and Medicine* **110**, 164–174 (2019).

25. M.-A. Carbonneau, V. Cheplygina, E. Granger, G. Gagnon, Multiple instance learning: A survey of problem characteristics and applications. *Pattern Recognition* **77**, 329–353 (2018).

26. H. Liu, W.-D. Xu, Z.-H. Shang, X.-D. Wang, K.-S. Wang, Breast Cancer Molecular Subtype Prediction on Pathological Images with Discriminative Patch Selection and Multi-Instance Learning. *Front. Oncol.* **12** (2022).

27. M. Bilous, M. Dowsett, W. Hanna, J. Isola, A. Lebeau, A. Moreno, F. Penault-Llorca, J. Rüschoff, G. Tomasic, M. van de Vijver, Current Perspectives on HER2 Testing: A Review of National Testing Guidelines. *Modern Pathology* **16**, 173–182 (2003).

28. E. A. Perez, J. Cortés, A. M. Gonzalez-Angulo, J. M. S. Bartlett, HER2 testing: Current status and future directions. *Cancer Treatment Reviews* **40**, 276–284 (2014).

29. A. I. Fernandez, M. Liu, A. Bellizzi, J. Brock, O. Fadare, K. Hanley, M. Harigopal, J. M. Jorns, M. G. Kuba, A. Ly, M. Podoll, K. Rabe, M. A. Sanders, K. Singh, O. L. Snir, T. R. Soong, S. Wei, H. Wen, S. Wong, E. Yoon, L. Pusztai, E. Reisenbichler, D. L. Rimm, Examination of Low ERBB2 Protein Expression in Breast Cancer Tissue. *JAMA Oncol* **8**, 1–4 (2022).

30. C. G. Johnson, J. C. Levenkron, A. L. Suchman, R. Manchester, Does physician uncertainty affect patient satisfaction? *Journal of general internal medicine* **3**, 144–149 (1988).

31. A. N. Meyer, T. D. Giardina, L. Khawaja, H. Singh, Patient and clinician experiences of uncertainty in the diagnostic process: current understanding and future directions. *Patient Education and Counseling* **104**, 2606–2615 (2021).

32. TissueArray.Com, *TissueArray.Com*. http://www.tissuearray.com/.

33. H. Yuen, J. Princen, J. Illingworth, J. Kittler, Comparative study of Hough Transform methods for circle finding. *Image and Vision Computing* **8**, 71–77 (1990).

34. G. Huang, Z. Liu, L. van der Maaten, K. Q. Weinberger, "Densely Connected Convolutional Networks" (2017; https://openaccess.thecvf.com/content_cvpr_2017/html/Huang_Densely_Connected_Convolutional_CVPR_2017_paper.html), pp. 4700–4708.

35. I. Loshchilov, F. Hutter, Decoupled Weight Decay Regularization. arXiv arXiv:1711.05101 [Preprint] (2019). http://arxiv.org/abs/1711.05101.